\documentclass[letterpaper,10pt]{article}

\usepackage{amsmath}
\usepackage{amsfonts}
\usepackage{amssymb}
\usepackage{latexsym}

\newcommand{\mathscr}[1]{{\bf #1}}

\begin{document}
\title{The Transplanckian Question and the Casimir Effect}
\author{Sven Bachmann${}^{a,d}$, ~~ Achim Kempf${}^{a,b,c}$ \\\\
Depts. of Applied Mathematics${}^{(a)}$ and Physics${}^{(b)}$,
University of Waterloo\\
Perimeter Institute for Theoretical Physics${}^{(c)}$\\
Waterloo, Ontario, Canada\\\\
Ecole Polytechnique F\'ed\'erale de Lausanne${}^{(d)}$
 \\ Lausanne, Switzerland}

\date{ }
\maketitle
\begin{abstract}
It is known that, through inflation, Planck scale phenomena should have
left an imprint in the cosmic microwave background. The magnitude of this
imprint is expected to be suppressed by a factor $\sigma^n$ where
$\sigma\approx 10^{-5}$ is the ratio of the Planck length to the Hubble
length during inflation. While there is no consensus about the value of
$n$, it is generally thought that $n$ will determine whether the imprint
is observable. Here, we suggest that the magnitude of the imprint may not
be suppressed by any power of $\sigma$ and that, instead, $\sigma$ may
merely quantify the amount of fine tuning required to achieve an imprint
of order one. To this end, we show that the UV/IR scale separation,
$\sigma$, in the analogous case of the Casimir effect plays exactly this
role.
\end{abstract}

\section{Introduction}

The so-called transplanckian question is concerned with low energy
phenomena whose calculation appears to require the validity of standard
quantum field theory (QFT) at energies beyond the Planck scale. The issue
first arose in the context of black holes: the derivation of Hawking
radiation is based on the assumption that standard QFT is valid even at
scales beyond the Planck scale. For example, the typical low-energy
Hawking photons that an observer might detect far from the horizon are
implied to have possessed proper frequencies that were much larger than
the Planck frequency close to the event horizon, even at distances from
the horizon that are farther than a Planck length. This led to the
question if Planck scale effects could influence or even invalidate the
prediction of Hawking radiation. Numerous studies have investigated the
issue and the current consensus is that Hawking radiation is largely
robust against modifying QFT in the ultraviolet (UV). This is plausible
since general thermodynamic considerations already constrain key
properties of Hawking radiation. See, e.g.,
\cite{brout-review-etc,Unruh2}.

More recently, the transplanckian question arose in the context of
inflationary cosmology: according to most inflationary models, space-time
inflated to the extent that fluctuations which are presently of
cosmological size started out with wavelengths that were shorter than the
Planck length. The derivation of the inflationary perturbation spectrum
therefore assumes the validity of standard QFT beyond the Planck scale.
Unlike in the case of black holes, no known thermodynamic reasons
constrain the properties of the inflationary perturbation spectrum so as
to make it robust against the influence of physics at the Planck scale. It
is, therefore, very actively being investigated if future precision
measurements of the cosmic microwave background (CMB) intensity and
polarization spectra could in this way offer an experimental window to
Planck scale phenomena. See e.g. \cite{infl-etc}.

It is generally expected that the imprint of Planck scale physics on the
CMB is suppressed by a factor $\sigma^n$ where $\sigma$ is defined as the
ratio of the UV and IR scale. In inflation, this ratio is $\sigma \approx
10^{-5}$ since modes evolve nontrivially only from the Planck scale to the
Hubble scale, $L_\text{Hubble}\approx 10^5 ~L_\text{Planck}$, after which
their dynamics freezes until much later when they reenter the horizon to
seed structure formation. We note that if the UV scale is the string
scale, $\sigma$ could be as large as $\sigma\approx 10^{-3}$. Regarding
the value of the power, $n$, in $\sigma^n$, no consensus has been reached.
It is generally expected however, that the value of $n$ decides whether
the imprint of Planck scale physics in the CMB could ever become
measurable.

Concrete studies in this field often model the influence of Planck scale
physics on QFT through dispersion relations that become nonlinear at high
energies. This approach is motivated by the fact that the natural
ultraviolet cutoff in condensed matter systems characteristically affects
the dispersion relations there. See, e.g., \cite{Unruh2}. It has been
shown that while some ultraviolet-modified dispersion relations would
affect the inflationary predictions for the CMB to the extent that effects
might become measurable, other modified dispersion relations would have a
negligible effect on the CMB. It is so far not fully understood which
properties of Planck scale modifications to the dispersion relation decide
whether or not an observable effect is induced. In order to clarify if and
how an imprint of Planck scale effects in the CMB are suppressed by
$\sigma$ it would be most interesting, therefore, to find and study the
operator which maps arbitrary ultraviolet-modified dispersion relations
directly into the correspondingly modified CMB perturbation spectra.

Here, we will investigate the simpler transplanckian question for the
Casimir force. As is well-known, the Casimir force arises due to quantum
fluctuations of the electromagnetic field and occurs between neutral
conducting objects. Similar to Hawking radiation and inflationary
fluctuations, the Casimir force can be seen as a vacuum effect which
involves modes of arbitrarily short wave lengths. In fact, naively it
appears that modes contribute the more the shorter their wave length is.
This suggests that, in principle, the predicted Casimir force could be
influenced by Planck scale physics.

The Casimir effect is simple enough so that we will be able to completely
answer its transplanckian question when modelling Planck scale physics
through ultraviolet-modified dispersion relations. Namely, we will find
the explicit operator which maps generic ultraviolet-modified dispersion
relations into the corresponding Casimir force functions. The properties
of this operator reveal that and how ultraviolet-modified dispersion
relations can strongly affect the Casimir force even in the `infrared'
i.e. at practically measurable distances. Interestingly, the extreme ratio
$\sigma\approx 10^{-28}$ between the effective UV and IR scales in the
Casimir effect does not suppress the possible strength of Planck scale
effects in the Casimir force at macroscopic distances. We find that,
instead, the extreme value of $\sigma$ implies that UV-modified dispersion
relations that lead to a large IR effect merely need to be extremely
fine-tuned, which suppresses the a priori likelihood that such a
dispersion relation should arise from an underlying theory of quantum
gravity. This is of interest because if the situation in inflation is
analogous, the imprint of Planck space physics in the CMB may not be
suppressed in strength by any power $\sigma^n$ of $\sigma$. Instead, the
$\sigma$ of inflation, $\sigma\approx 10^{-5}$ or $\sigma\approx10^{-3}$,
may determine the amount of fine-tuning required to achieve an imprint of
order one. Thus, $\sigma$ would be related to the a priori likelihood for
an observable imprint to arise from an underlying theory of quantum
gravity. In inflation, this likelihood would not be extremely small since
the UV and IR scales in inflation are not extremely separated.

\section{The Casimir force and ultraviolet-modified dispersion relations}

The Casimir effect arises when reflecting surfaces pose boundary
conditions on the modes of the electromagnetic field. For example, two
perfectly reflecting parallel plates impose boundary conditions such that
the set of electromagnetic modes in between them is discretized. The
spacing of the modes, and therefore the vacuum energy that each mode
contributes, depends on the distance between the plates. This
distance-dependence of the vacuum energy leads to the Casimir force
between the plates. In general, the force is a function of both the
distance and the shape of the reflecting surfaces, and the force can be
both attractive or repulsive.

The Casimir effect was first predicted, by Casimir, in 1948, see
\cite{Casimir:1948dh}. In the meanwhile, the Casimir force has been
calculated for several types of geometries and in various dimensions.
Also, effects of imperfect conductors, rough surfaces and finite
temperatures have been considered, see \cite{Balian:2002}. In addition,
detailed calculations have been carried out to account for higher order
corrections due to virtual electrons and their interaction with the
boundaries \cite{Aghababaie:2003iw}. For recent reviews see
\cite{bordag-etal} and for precision measurements of the effect see e.g.
\cite{Lamoreaux:1999cu-etal}.

For our purposes, the essential features of the Casimir effect are
captured already when working with a massless real scalar field between
two perfectly conducting parallel plates. For simplicity, we will consider
the simple case of just one space dimension, in which case the reflecting
plates are mere points. We place these points at $x=0$ and $x=L$, i.e., we
impose the boundary conditions $\hat{\phi}(0,t)=0=\hat{\phi}(L,t)$ for all
$t$. In order to fulfill these boundary conditions we expand the quantum
field between the plates using the Fourier sine series:
\begin{equation}
\hat{\phi}(x,t) = \sum_{n=1}^{\infty} \hat{\phi}_n(t) \sin(k_n x), ~~~~~~~
k_n= \frac{n\pi}{L}
\end{equation}
We are using units such that $\hbar=c=1$. Recall that in a Fourier sine
series all $n$ and therefore all wave numbers $k_n$ are positive. The
reason is that the sine functions form a complete eigenbasis of the square
of the momentum operator, $\hat{p}^2=-d^2/dx^2$, all of whose eigenvalues
are of course positive. (Recall that the momentum operator of a particle
in a box is not self-adjoint and not diagonalizable, see e.g.
\cite{ak-beethoven}). The usual ansatz
\begin{equation}
\hat{\phi}_n=\frac{1}{\sqrt{\omega(k_n)L}}\left(e^{i\omega(k_n)t}a^\dagger_n+
e^{-i\omega(k_n)t}a_n\right)
\end{equation}
with $[a_n,a_m^\dagger]=\delta_{n,m}$ diagonalizes the Hamiltonian:
\begin{equation}
\hat{H}=\sum_{n=1}^\infty
\omega(k_n)\left(a^\dagger_na_n+\frac{1}{2}\right)
\end{equation}
Thus, with the usual linear dispersion relation
\begin{equation}
\omega(k)=k,
\end{equation}
the vacuum energy between plates of distance $L$ is divergent:
\begin{eqnarray}
\label{eq:infSum} E_{in}(L) & = &
\frac{1}{2}\sum_{n=0}^{\infty}\omega(k_{n}) \\
& = & \frac{\pi}{2L}\sum_{n=0}^{\infty} n ~~~=~ \infty
\end{eqnarray}
We notice that modes appear to contribute the more the shorter their
wavelength, i.e. the larger $k$ and $n$ are. One proceeds by regularizing
the divergence and by then calculating the \it change \rm in the
regularized total energy (of a large region that contains the plates) when
varying $L$. As is well-known, the resulting expression for the Casimir
force remains finite after the regularization is removed, and reads:
\begin{equation}
\mathcal{F}(L)=-\frac{\pi}{24L^{2}}\,
\end{equation}
It has been shown that this result does not depend on the choice of
regularization method. Our aim now is to re-calculate the Casimir force
within standard quantum field theory while modelling the onset of Planck
scale phenomena at high energies through general nonlinear modifications
to the dispersion relation. The goal is to calculate the operator which
maps arbitrary modified dispersion relations $\omega(k)$ into the
resulting Casimir force functions $\mathcal{F}(L)$. To this end, let us
begin by writing generalized dispersion relations in the form:
\begin{equation}
\omega(k)=k_{c}f\left(\frac{k}{k_{c}}\right)
\end{equation}
Here, $k_{c}>0$ is a constant with the units of momentum, say the Planck
momentum so that its inverse is the Planck length: $L_c=k_c^{-1}$. The
function $f$ encodes unknown Planck scale physics and for now we will make
only these minimal assumptions:
\begin{itemize}
\label{ma}
\item $f(0)=0$, and $f(x)\approx x$ if $x\ll 1$
~~~(regular dispersion at low energies)
\item $f(x)\geq 0$ when $x
\ge0$ ~~~~(stability: each mode carries positive energy)
\end{itemize}
We will use the term dispersion relation for both $\omega(k)$ and $f(x)$.
\label{bedi}
\section{Exponential regularization}

For generically modified dispersion relations the vacuum energy
(\ref{eq:infSum}) must be assumed to be divergent and therefore in need of
regularization. Let us therefore regularize (\ref{eq:infSum}) by
introducing an exponential regularization function, parametrized by
$\alpha>0$, i.e. we define the regularized vacuum energy between the
plates as:
\begin{equation}
\label{eq:EinReg} E_{in}^{reg}(L)=\frac{1}{2}\sum_{n=0}^{\infty}k_{c}
\,f\left(\frac{n\pi}{k_{c}L}\right)\exp{\left[-\alpha
k_{c}\,f\left(\frac{n\pi}{k_{c}L}\right)\right]}\,
\end{equation}
In order to calculate the regularized vacuum energy density outside the
plates we notice that the right and left outside regions are half axes and
that the energy density in a half axis can be calculated from
(\ref{eq:EinReg}) by letting $L$ go to infinity:
\begin{equation}
\mathcal{E}^{reg}=\lim_{L\to\infty}\frac{E_{in}^{reg}(L)}{L}\, \label{ove}
\end{equation}
The expression for the vacuum energy density outside the plates,
(\ref{ove}), is conveniently rewritten as a Riemann sum by defining
$\Delta x = \frac{1}{L}$:
\begin{eqnarray}
\label{eq:EoutReg} \mathcal{E}^{reg} &
 = & \lim_{\Delta x\to
0}\left\{\frac{1}{2}\sum_{n=0}^{\infty}\Delta
x~k_{c}\,f\left(\frac{n\Delta x\pi}{k_{c}}\right)\exp{\left[-\alpha
k_{c}\,f\left(\frac{n\Delta
x\pi}{k_{c}}\right)\right]} \right\}\nonumber \\
& = &\frac{k_{c}^{2}}{2\pi}\int_{0}^{\infty}dx\,f(x) \exp{\left[-\alpha
k_{c}\,f(x)\right]}\,.
\end{eqnarray}
Notice that we are here implicitly restricting attention to dispersion
relations for which exponential regularization is sufficient to render the
energy densities outside and between the plates finite. This excludes, for
example, the dispersion relation $f(x)=\ln(1+x)$ which would require a
regularization function such as $\exp(-f(x)^2)$. We will later be able to
lift this restriction on the dispersion relations, namely by allowing the
use of arbitrary regularization functions. Indeed, as we will prove in
Sec.\ref{indep}, our results only depend on the dispersion relation and
are independent of the choice of regularization function, as long as the
regularization function does regularize the occurring series and
integrals, obeys certain mild smoothness conditions and recovers the
original divergent series of (\ref{eq:infSum}) in the limit $\alpha\to 0$.

In order to calculate the Casimir force, let us now consider a very large
but finite region, say of length $M$, which contains the two plates. The
total energy in this region is finite and consists of the energy between
the plates, (\ref{eq:EinReg}), plus the energy density outside the plates,
(\ref{eq:EoutReg}), multiplied by the size of the region outside, namely
$M-L$. Note that by choosing $M$ large enough ensures that the energy
density outside the plates does not depend on $L$. Thus, the total energy
in this region is given by $E_{in}^{reg}(L)+(M-L)\mathcal{E}^{reg}$. The
regularized Casimir force is the derivative of this energy with respect to
a change in the distance of the plates:
\begin{equation}
\mathcal{F}_{\alpha}(L)=-\frac{\partial}{\partial L}E_{in}^{reg}+
\mathcal{E}^{reg}\,.
\end{equation}
The total length $M$ of the region under consideration has dropped out, as
it should be. Hence, before removing the regularization (i.e. before
letting $\alpha \rightarrow 0^+$), the Casimir force in the presence of a
nonlinear dispersion relation is given by:
\begin{eqnarray}
\label{eq:GenRel} \lefteqn{\mathcal{F}_{\alpha}(L)=\frac{1}{2}k_{c}\left\{
\sum_{n=0}^{\infty}\frac{1}{L}\left[ \frac{n\pi}{k_{c}L}~ f^{\prime}
\left(\frac{n\pi}{k_{c}L}\right)\exp{\left[-\alpha k_{c}\,f\left(
\frac{n\pi}{k_{c}L}\right)\right]}\times{}
\right.\right.}\nonumber \\
& & {}\left.\left. \times \left(1-\alpha k_{c}\,f
\left(\frac{n\pi}{k_{c}L}\right)\right)\right]+
\frac{k_{c}}{\pi}\int_{0}^{\infty}dx\,f(x)\exp{ \left[-\alpha
k_{c}\,f(x)\right]}\right\}
\end{eqnarray}
Here, $f'$ stands for differentiating $f$ with respect to the variable
$x=\frac{n\pi}{k_cL}$.
\section{Application of the Euler-Maclaurin formula}
It will be convenient to collect the terms that constitute the argument of
the series in a new definition:
\begin{equation}
\label{tfg} \varphi_{\alpha}(t) := \frac{t\pi}{k_{c}L} ~f^{\prime}
\left(\frac{t\pi}{k_{c}L}\right)\exp{\left[-\alpha
k_{c}\,f\left(\frac{t\pi}{k_{c}L}\right)\right]}\left(1-\alpha
k_{c}\,f\left(\frac{t\pi}{k_{c}L}\right)\right)
\end{equation}
Thus, (\ref{eq:GenRel}) becomes:
\begin{equation}
\mathcal{F}_{\alpha}(L)=\frac{k_c}{2L} \sum_{n=0}^{\infty}
\varphi_\alpha(n) ~+~ \frac{k_c^2}{2\pi}\int_0^\infty dx~f(x)~e^{-\alpha
k_c f(x)}\label{cf45}
\end{equation}
We notice that if the first term in (\ref{cf45}) were an integral instead
of a series then the two terms in (\ref{cf45}) would exactly cancel
another:
\begin{eqnarray}
\label{eq:intPrts}\label{3a}
\frac{k_c}{2L}\int_{0}^{\infty}\varphi_{\alpha}(t)\,dt & = &
\frac{k_c}{2L}\frac{k_cL}{\pi}\int_{0}^{\infty}\varphi_{\alpha}(t)
\,\frac{\pi}{k_cL}\,dt \\
& = & \frac{k_c^2}{2\pi}\int_{0}^{\infty}dx \,x\,f^{\prime}(x)e^{-\alpha
k_{c}f(x)}\left(1-\alpha
k_{c}f(x)\right)\\
& = & \left .  \frac{k_c^2}{2\pi}~xf(x)e^{-\alpha
k_{c}f(x)}\right|_{0}^{\infty}-\frac{k_c^2}{2\pi}\int_{0}^{\infty}dx
\,f(x)e^{-\alpha k_{c}f(x)}\label{bt}\\ \label{3c} \label{hew} & = &
0-\frac{k_c^2}{2\pi}\int_{0}^{\infty}dx\,f(x)e^{-\alpha k_{c}f(x)}\,.
\end{eqnarray}
In (\ref{bt}), the boundary terms are zero because at $x=0$ the dispersion
relation yields $f(0)=0$ and because for $x\rightarrow \infty$ the
finiteness of (\ref{eq:EoutReg}) implies that its integrand decays faster
than $1/x$.

In order to compute the Casimir force, let us now use the Euler-Maclaurin
sum formula, see e.g. \cite{Ford-etal}, to express the series of
$\varphi_\alpha$ as an integral of $\varphi_\alpha$ plus corrections. As
we just saw, the integral will then cancel in (\ref{cf45}) and the
correction terms will constitute the Casimir force. To this end, recall
that if the $(k+1)$st derivative of a function $\xi$ is continuous, i.e.,
if $\xi\in \mathcal{C}^{k+1}$, then:
\begin{eqnarray}
\label{eq:em1} \sum_{a<n\le b}\xi(n)&=&\int_a^b \xi(t)\,dt + \sum_{r=0}^k
\frac{(-1)^{r+1}B_{r+1}}{(r+1)!}
\left(\xi^{(r)}(b)-\xi^{(r)}(a)\right) +\nonumber \\
& & +\frac{(-1)^k}{(k+1)!} \int_a^b B_{k+1}(t)\xi^{(k+1)}(t)\,dt
\end{eqnarray}
Here, the superscript at $\xi^{(r)}$ denotes the $r$'th derivative of the
function $\xi$, the $B_{s}$ are the Bernoulli numbers and $B_{s}(t)$ is
the $s$'th Bernoulli periodic function, i.e. the periodic extension of the
$s$'th Bernoulli polynomial from the interval $[0,1]$.

We can now choose $\xi=\varphi_{\alpha}$, set $a=0$ and take the limit
$b\to \infty$. Since the vacuum energy density, (\ref{eq:EoutReg}), is
finite it follows that (\ref{3c}) is finite and therefore also (\ref{3a}).
This in turn implies that $\lim_{x\rightarrow \infty} \varphi_{\alpha}(x)=
0$ and $\lim_{x\to\infty}\varphi_{\alpha}^{(n)}(x)=0$ for all $n\geq1$.
Hence, the series involving the Bernoulli numbers simplifies and we obtain
for arbitrary $k \in \mathbb{N}$ this Euler-Maclaurin formula for
$\varphi_\alpha$:
\begin{equation}
\sum_{n=0}^{\infty}\varphi_{\alpha}(n)= \int_{0}^{\infty}
\varphi_{\alpha}(t)\,dt-\sum_{r=0}^{k}
\frac{(-1)^{r+1}B_{r+1}}{(r+1)!}~\varphi_{
\alpha}^{(r)}(0)+\Omega_k[\varphi_{\alpha}]\,
\end{equation}
Here, $\Omega_k[\varphi_{\alpha}]$ represents the remainder integral:
\begin{equation}
\Omega_k[\varphi_{\alpha}] = \frac{(-1)^k}{(k+1)!}\int_0^\infty
B_{k+1}(t)~ \varphi_\alpha^{(k+1)}(t)~dt \label{rema}
\end{equation}
Using $\varphi_{\alpha}(0)=0$ and the fact that, except for $B_{1}$, all
Bernoulli numbers $B_s$ with odd indices $s$ are zero, we obtain:
\begin{equation}
\sum_{n=0}^{\infty}\varphi_{\alpha}(n)= \int_{0}^{
\infty}\varphi_{\alpha}(t)\,dt-\sum_{r=1}^{k} \frac{B_{2r}}{(2r)!}~
\varphi_{\alpha}^{(2r-1)}(0)+\Omega_k[\varphi_{\alpha}]\label{drt6}
\end{equation}
Equation (\ref{drt6}) expresses the series as an integral plus
corrections, as desired. Applied to the expression (\ref{cf45}) for the
regularized Casimir force, $\mathcal{F}_\alpha(L)$, the integrals then
cancel and we obtain for the regularized Casimir force:
\begin{equation}
\label{eq:Falpha}
\mathcal{F}_{\alpha}(L)=-\frac{k_{c}}{2L}\sum_{r=1}^{k}\frac{B_{2r}}{2r!}
 ~ \varphi_{\alpha}^{(2r-1)}(0)+\frac{k_c}{2L}~
 \Omega_k[\varphi_{\alpha}]
\end{equation}
The actual Casimir force, $\mathcal{F}(L)$, is obtained by removing the
regularization:
\begin{equation}
\mathcal{F}(L)=\lim_{\alpha\to 0^+}\left\{
-\frac{k_{c}}{2L}\sum_{r=1}^{k}\frac{B_{2r}}{2r!}
 ~ \varphi_{\alpha}^{(2r-1)}(0)+\frac{k_c}{2L}~
 \Omega_k[\varphi_{\alpha}]\right\}\label{cf45b}
\end{equation}
\section{The Casimir force for polynomial dispersion relations}

In order to further evaluate this expression for the Casimir force let us
restrict attention to dispersion relations that are sufficiently well
behaved so that $\varphi_\alpha(t)$ is $\mathcal{C}^\infty$ with respect
to both $\alpha$ and $t$. The simplest case is that of dispersion
relations which are polynomial:
\begin{equation}
f(x)=\sum_{s=0}^{n}\nu_{s}x^{s} \label{polone}
\end{equation}
We are assuming that $\varphi_\alpha(t) \in \mathcal{C}^\infty$ which here
allows us to take the limit $\alpha \to 0$ in $\varphi_\alpha(t)$ before
differentiating it. From (\ref{tfg}) we then have $\varphi_{0}(t) =
\lim_{\alpha\to 0}\varphi_{\alpha}(t)=x(t) f^{\prime}(x(t))$ where
$x(t)=\frac{t\pi}{k_cL}$ and where $'$ stands for $d/dx$. Thus, iterated
differentiation yields
\begin{equation}
\label{eq:diffPhi} \frac{d^{n}\varphi_{0}(t)}{dt^{n}} = n
\left(\frac{\pi}{k_cL}\right)^n
\,\frac{d^nf(x)}{dx^n}+x\left(\frac{\pi}{k_cL}\right)^{n+1}\,
\frac{d^{n+1}f(x)}{dx^{n+1}}
\end{equation}
and therefore the terms in the series in (\ref{cf45b}) read:
\begin{equation}
\varphi^{(n)}_{0}(t)\vert_{t=0}= n\left(\frac{\pi}{k_cL}\right)^n
\,f^{(n)}(x)\vert_{x=0}\label{dfg}
\end{equation}
We now show that the remainder term $\Omega_k[\varphi_\alpha]$ does not
contribute. Assuming for the moment that the dispersion relation is
polynomial, $\varphi_{\alpha}(t)$ is a polynomial times the exponential
regularization function $e^{-\alpha k_c f}$ which tends to $1$ as
$\alpha\to 0$. Therefore, after sufficiently many differentiations, i.e.,
when choosing $k$ large enough, $\varphi^{(k+1)}_{\alpha}(t)\to 0$ as
$\alpha\to 0$ for all fixed $t$. In order to evaluate
$\Omega_k[\varphi_\alpha]$, let us now split (\ref{rema}) into two
integrals: $\int_0^\infty=\int_0^b+\int_b^\infty$. For all finite $b>0$
the first integral commutes with the limit $\alpha \to 0$ to yield for
large enough $k$:
\begin{equation}
\lim_{\alpha\to 0}\int_0^b
B_{k+1}(t)~\varphi_\alpha^{(k+1)}(t)~dt=\int_0^b\lim_{\alpha\to 0}
B_{k+1}(t)~\varphi_\alpha^{(k+1)}(t)~dt=0
\end{equation}
Further, we notice that, since $f$ is polynomial and the exponential
regularization function is positive, $\varphi_\alpha(t)$ does not change
sign for all $t>b$ if $b$ is chosen sufficiently large. Since the periodic
Bernoulli functions are bounded from above by their Bernoulli numbers we
therefore obtain:
\begin{eqnarray}
\left\vert\int_b^\infty B_{k+1}(t)~\varphi_\alpha^{(k+1)}(t)~dt\right\vert
 & \le & \left\vert B_{k+1}\right\vert~
\left\vert\int_b^\infty \varphi_\alpha^{(k+1)}(t)~dt\right\vert\nonumber\\
 & \le & \left\vert
 B_{k+1}\right\vert~\left\vert\varphi_\alpha^{(k)}(t)
 \vert_b^\infty\right\vert\nonumber\\
& = & \left\vert
 B_{k+1}\right\vert~\left\vert\varphi_\alpha^{(k)}(b)
 \right\vert\nonumber\\
 & \to & 0 ~~~\mbox{as}~~\alpha \to 0
\end{eqnarray}
Thus, when choosing $k$ large enough, the remainder term disappears so
that, using (\ref{dfg}), we obtain for the Casimir force for arbitrary
polynomial dispersion relations:
\begin{equation} \label{eq:force}
\mathcal{F}(L)=-\frac{k_{c}}{2L}\sum_{r=1}^{k}
\frac{(2r-1)B_{2r}}{2r!}\,f^{(2r-1)}(0)\left(\frac{\pi}{
k_{c}L}\right)^{2r-1}
\end{equation}
Further, since $f^{(s)}(0)=s!\,\nu_{s}$, we obtain:
\begin{equation}
\label{eq:forcePoly} \mathcal{F}(L) = -\frac{k_{c}}{2L}\sum_{r=1}^k
\frac{(2r-1)B_{2r}}{2r}~\nu_{2r-1} \left(\frac{\pi}{k_{c}L}\right)^{2r-1}
\end{equation}
We notice that, interestingly, the even powers in a nonlinear dispersion
relation, i.e. the coefficients $\nu_{2r}$, do not contribute to the
Casimir force.
\smallskip\newline
As a consistency check, let us now choose the usual linear dispersion
relation $f(x)=x$. Since $B_{2}=\frac{1}{6}$, we obtain
\begin{equation}
\mathcal{F}(L)=-\left(\frac{k_{c}}{2L}\right)
\left(\frac{\pi}{k_{c}L}\right)\frac{1}{2\cdot 6}=-\frac{\pi}{24L^{2}}\,,
\end{equation}
which is the well-known usual result for the Casimir force, as it should
be.
\section{Generic dispersion relations}
Considering our results for the Casimir force with polynomial dispersion
relations, (\ref{eq:force},\ref{eq:forcePoly}) we notice that the addition
of mode energies translates into the addition of the corresponding Casimir
forces: if two dispersion relations are added, $f_{t}(x)=f_1(x)+f_2(x)$,
then the two corresponding Casimir forces are added:
\begin{equation}
\label{eq:additivity} \mathcal{F}_{t}=\mathcal{F}_1+\mathcal{F}_2
\end{equation}
This shows that the operator, $\mathcal{K}$, that we have been looking
for, namely the operator which maps arbitrary dispersion relations into
their corresponding Casimir forces, $\mathcal{K}: f \mapsto \mathcal{F}$,
is a linear operator:
\begin{equation}
\mathcal{K}[f_1+f_2]=\mathcal{K}[f_1]+\mathcal{K}[f_2]
\end{equation}
Because of its linearity, we can straightforwardly extend the action of
$\mathcal{K}$ to arbitrary dispersion relation, $f$, which are given by
power series in $x$:
\begin{equation}
f(x)=\sum_{s=0}^{\infty}\nu_{s}x^{s}
\end{equation}
The radius of convergence of the power series must be infinite since the
dispersion relation needs to be evaluated for all $x$, i.e., $f$ is an
entire function. The linearity of $\mathcal{K}$ yields the corresponding
Casimir force function $\mathcal{F}$ as a power series in $1/L$:
\begin{equation}
\label{foww}
\mathcal{K}[f](L)=\mathcal{F}(L)=-\frac{k_{c}}{2L}\sum_{r=1}^{\infty}
\frac{(2r-1)B_{2r}}{2r}\,\nu_{2r-1}\left(\frac{\pi}{
k_{c}L}\right)^{2r-1}\,
\end{equation}
We need to determine under which conditions the resulting power series for
the Casimir force function is convergent. Interestingly, as we will show
in Sec.\ref{nese}, the convergence, i.e. the well-definedness of the
Casimir force, generally depends on the plate separation $L$. When the
power series possesses a finite radius of convergence, i.e. when there is
a largest allowed value for $1/L$, this means that there is a smallest
allowed value for the length $L$. This is beautifully consistent with the
expectation that dispersion relations that arise from an underlying
quantum gravity theory can imply a finite minimum length scale.

For analyzing the convergence properties of the series (\ref{foww}) the
presence of the Bernoulli numbers is somewhat cumbersome. It will be
useful, therefore, to use the connection between the Bernoulli numbers and
the Riemann zeta function, see \cite{Havil}:
\begin{equation}
B_{n}=(-1)^{n+1}n\,\zeta(1-n)
\end{equation}
Thus:
\begin{equation}
\label{mres} \mathcal{F}(L)=\frac{k_{c}}{2L}\sum_{r=1}^{\infty}
(2r-1)\zeta(1-2r)~\nu_{2r-1}\left(\frac{ \pi}{k_{c}L}\right)^{2r-1}\,.
\end{equation}
We can now use the fact that, see \cite{Hardy99}:
\begin{equation}
\zeta(1-s)=\frac{2}{(2\pi)^{s}}\cos\left(\frac{1}{2}\pi
s\right)\Gamma(s)\zeta(s)
\end{equation}
In our case, since $s$ is always an integer, the Euler gamma function
reduces to a factorial, and the cosine is $\pm 1$. Thus:
\begin{equation}
\label{eq:forceFactorial}
\mathcal{F}(L)=\frac{k_{c}}{L}\sum_{r=1}^{\infty}
\frac{(-1)^{r}}{(2\pi)^{2r}}~(2r-1)~(2r-1)!~\zeta(2r)~
\nu_{2r-1}\left(\frac{\pi}{k_{c}L} \right)^{2r-1}
\end{equation}
Having replaced the Bernoulli numbers by the Riemann zeta function is
advantageous because obviously $\zeta(r)\to1$ very quickly as
$r\to\infty$. For example, for $r=6$, the difference is already at the one
percent level. This means that for the purpose of analyzing the
convergence properties of the power series we will be able to use that the
Riemann zeta function for the arguments that occur is close to $1$ and
essentially constant.

\section{Example with minimum length}
\label{nese} 
\begin{figure}[!ht]
\begin{center}
\input{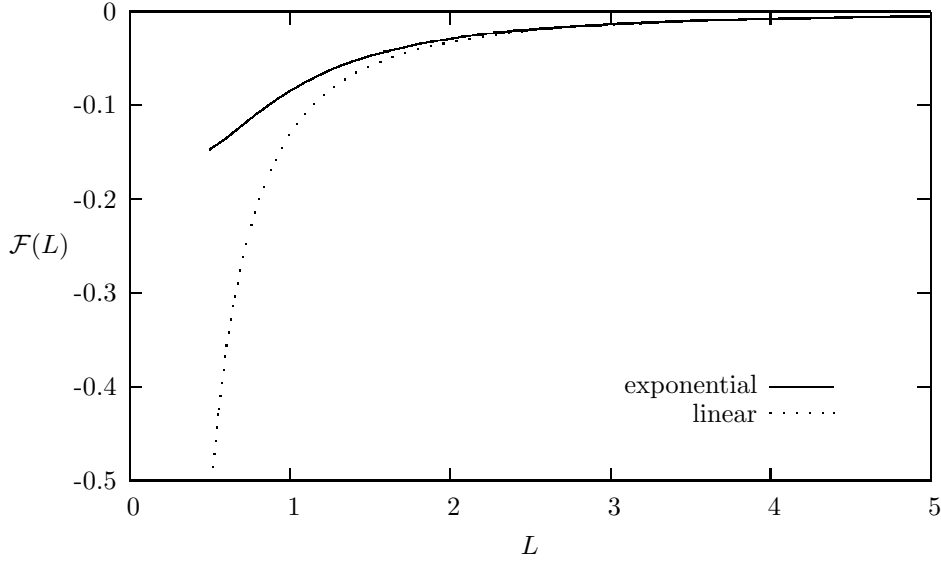}
\caption{The Casimir force for the exponential dispersion relation
$\omega(k)=k_c(\exp(k/k_c)-1)$. Note that the Casimir force is defined
only for $L$ larger than the finite minimum length $L_{min}=1/2$ (in units
of $1/k_c$).} \label{fig:exponential}
\end{center}
\end{figure}

Ultraviolet-modified nonlinear dispersion relations which
approach the usual linear dispersion relation for small momenta are given,
for example, by:
\begin{equation}
\label{eq:dispExpSinh} f(x)=\exp(x)-1 \qquad\textrm{and}\qquad
f(x)=\sinh(x)
\end{equation}
The odd coefficients, $\nu_{2r-1}=1/(2r-1)!$ are the same for both the
exponential and the $sinh$ dispersion relation, i.e. the two functions
differ only by their even part. But we know from (\ref{eq:forceFactorial})
that the even components of the dispersion relations do not affect the
Casimir force. The two dispersion relations therefore happen to lead to
the same Casimir force. It is plotted with the usual Casimir force in
Fig.\ref{fig:exponential}.
\newline
We see that the Casimir force matches the usual Casimir force at large $L$
but is weaker for small $L$. As the plot also shows, the Casimir force is
well defined only for values of $L$ above a certain value $L_c$,
corresponding to a finite radius of convergence of the power series in
$1/L$ for the Casimir force. In order to calculate this minimum length
$L_c$, we notice that all the coefficients $\nu_{2r-1}$ are non-negative,
which implies that (\ref{eq:forceFactorial}) is an alternating series.
Such series converge if and only if their coefficients converge to zero.
Hence, for any such dispersion relation, the Casimir force is well defined
for all $L$ which obey:
\begin{equation}
\lim_{r\to\infty}\left[\frac{1}{(2\pi)^{2r}}~
(2r-1)~(2r-1)!~\zeta(2r)~\nu_{2r-1}\left(\frac{\pi}{k_{c}L}
\right)^{2r-1}\right]=0
\end{equation}
In the particular case of the two dispersion relations above, we have
$\nu_{2r-1}=1/(2r-1)!$ and the condition that the Casimir force be
well-defined therefore reads
\begin{equation}
\lim_{r\to\infty}\left[\frac{\zeta(2r)}{(2\pi)}~
(2r-1)\left(\frac{1}{2k_{c}L} \right)^{2r-1}\right]=0
\end{equation}
which means that $\frac{1}{2k_cL} < 1$. The minimum length implied by this
dispersion relation is therefore:
\begin{equation}
L_c = \frac{1}{2k_c} \label{edf}
\end{equation}
This is an example of what we hinted at before, namely that a dispersion
relation can in this way reveal an underlying short-distance cutoff.
\smallskip\newline
For general dispersion relations the coefficients $\nu_{2r-1}$ are not
necessarily all positive, i.e., the Casimir force need not be given by an
alternating series. In this general case the minimum length can be
determined by using the fact that the radius of convergence,
$\mathcal{R}$, of an arbitrary power series $\sum c_{r}x^{r}$ is given by:
\begin{equation}
\label{eq:RConv}
\frac{1}{\mathcal{R}}=\limsup_{r\to\infty}\left|c_{r}\right|^{\frac{1}{r}}\,.
\end{equation}
For example, in the case of the dispersion relations given in
(\ref{eq:dispExpSinh}), where $\nu_{2r-1}=1/(2r-1)!$, the Casimir force
(\ref{eq:forceFactorial}) can be written as a power series
$\mathcal{F}(L)=\sum_{r=1}^\infty c_r \left(\frac{1}{L^2}\right)^r$ in
$1/L^2$ with the coefficients:
\begin{equation}
c_r = \frac{(-1)^r~k_c ~(2r-1)~\zeta(2r)}{2\pi}\left(\frac{1}{2 k_c
}\right)^{2r-1}
\end{equation}
Thus, the minimum length obeys
\begin{eqnarray}
L_c^2 & = & \limsup_{r\to\infty}\left[\frac{k_c
~(2r-1)~\zeta(2r)}{2\pi}\left(\frac{1}{2 k_c}\right)^{2r-1}\right]^{\frac{1}{r}}\\
& = & \lim_{r\to \infty}\left(\frac{1}{2 k_c}\right)^{\frac{2r-1}{r}}\\
& = & \left(\frac{1}{2k_c}\right)^2
\end{eqnarray}
and therefore:
\begin{equation}
L>L_c=\frac{1}{2k_{c}}
\end{equation}
As expected, this agrees with the result (\ref{edf}) which we obtained by
using the alternating series test.

\section{Regularization-function independence}
\label{indep} It is known that the prediction for the Casimir force with
the usual linear dispersion relation does not depend on the choice of
regularization function, as long as the regularization function obeys
certain smoothness conditions and is such that it does in fact regularize
the integrals and series which occur in the calculation.

In our calculation of the Casimir force for nonlinear dispersion relations
we chose an exponential regularization function. We need to prove that our
result (\ref{eq:force}) does not depend on this choice. To see that this
indeed the case, assume that we use an arbitrary regularization function,
$\gamma_{\alpha}(x)$, which is a positive function of $x$ that obeys
$\lim_{\alpha\to 0^+}\gamma_{\alpha}(x)=1$ for all $x$ so that the
original divergent series is recovered when the regulator $\alpha$ goes to
zero. The regularized energy between the plates then reads:
\begin{equation}
\label{eds} \tilde{E}_{in}^{reg}=\frac{1}{2}
\sum_{n=0}^{\infty}k_{c}\,f\left(\frac{n\pi}{k_{c}L}
\right)\,\gamma_{\alpha}\left[f\left(\frac{n\pi}{k_{c}L} \right)\right]
\end{equation}
The regularization function, $\gamma_{\alpha}$, needs to be chosen such
that (\ref{eds}) as well as the energy density are finite, i.e. such that
$\lim_{L\to\infty} \tilde{E}_{in}^{reg}(L)/L<\infty$, which means:
\begin{equation}
\label{eq:finiteCut}
\int_{0}^{\infty}dx\,f(x)\gamma_{\alpha}\left[f(x)\right]<\infty\,
\end{equation}
Finally, in order to be able to use the Euler-Maclaurin sum formula and in
it to interchange $d/dt$ and the limit $\alpha\to 0$, we require the
regularization functions $\gamma_{\alpha}$ to be smooth enough so that
$\gamma_{\alpha}\in\mathcal{C}^{\infty}$ as well as $\varphi_\alpha(t) \in
\mathcal{C}^{\infty}$ as a function of $\alpha$ and $t$. The above
derivation of the Casimir force can then be repeated point by point using
the corresponding new definition of $\varphi_{\alpha}$. In particular, we
apply the Euler-Maclaurin sum formula to the expression:
\begin{eqnarray}
\label{eq:GenGenRel}
\lefteqn{\tilde{\mathcal{F}}_{\alpha}(L)=-\frac{k_{c}}{2L}\left\{
\sum_{n=0}^{\infty}\left[ \frac{n\pi}{k_{c}L}~
f^{\prime}\left(\frac{n\pi}{k_{c}L}\right)
\left\{\gamma_{\alpha}\left[f\left(\frac{n\pi}{k_{c}L}\right)\right]+{}
\right.\right.\right.}\nonumber \\
& & {}\left.\left.\left. +f\left(\frac{n\pi}{k_{c}L}\right)
\gamma_{\alpha}^{\prime} \left[f\left(\frac{n\pi}{k_{c}L}\right)\right]
\right\}\right]+\frac{k_{c}L}{\pi}\int_{0}^{
\infty}f(x)\gamma_{\alpha}[f(x)]\right\}
\end{eqnarray}
An integration by parts as in (\ref{eq:intPrts}) shows that the integrals
cancel. Equation (\ref{eq:finiteCut}) ensures that the boundary term
vanishes, as before in (\ref{bt}). Hence, we again arrive at
(\ref{eq:Falpha}). We now take the limit $\alpha\to 0$ term by term in the
sum, and since $\varphi_{\alpha}$ is in $\mathcal{C}^{\infty}$, we can
again do this before differentiating. Moreover, by the basic assumptions
made on $\gamma_{\alpha}$, we know that $\gamma_{\alpha}^{\prime}(x)\to 0$
as $\alpha\to 0$, so that as before:
\begin{equation}
\lim_{\alpha\to 0}\varphi_{\alpha}(t)=x(t)f^{\prime}(x(t))
\end{equation}
The arguments given in the previous section to show that the remainder
integral disappears for polynomial dispersion relations and that the
coefficients in the Euler-Maclaurin sum are those given in
(\ref{eq:forcePoly}) apply unchanged. This proves that our results for the
Casimir force are independent of the choice of regularization function, as
it should be.

\section{The operator $\mathcal{K}$ which maps dispersion
relations into Casimir force functions}

In preparation for our study of the transplanckian question for the
Casimir effect in Sec.\ref{tps}, let us now calculate explicit
representations of the operator $\mathcal{K}$ which maps dispersion
relations $f$ into Casimir force functions $\mathcal{F}$:
\begin{equation}
\mathcal{K}: ~f(x) \longmapsto \mathcal{F}(L)
\end{equation}
We already saw that $\mathcal{K}$ is linear. Indeed, from
(\ref{eq:forceFactorial}), it can be written as a differential operator:
\begin{equation}
\mathcal{K}=\frac{k_{c}}{2\pi
L}\sum_{r=1}^{\infty}(-1)^{r}(2r-1)\zeta(2r)\left(\frac{1}{2k_{c}L}
\right)^{(2r-1)}\left .\frac{d^{(2r-1)}}{dx^{(2r-1)}}\right|_{x=0}
\label{dop}
\end{equation}
As we already mentioned, the convergence of the zeta function,
$\zeta(2r)\to 1$, is very fast as $r\to\infty$. Since the study of the
transplanckian question involves large orders of magnitudes, we will
therefore henceforth replace $\zeta(2r)$ by $1$. By this approximation we
incur at most a numerical error of a pre-factor of order one which will
not affect our later analysis of the question when the ultraviolet
modifications to the dispersion relations can or cannot affect the Casimir
force in the infrared.

\subsection{Representation of $\mathcal{K}$ as an integral operator}
\label{sec9.1} For the purpose of studying the transplanckian question,
the representation of $\mathcal{K}$ as a differentiation operator in
(\ref{dop}) is not as suitable as a representation as an integral operator
would be. Indeed, as we now show, an equivalent representation of
$\mathcal{K}$ is given by
\begin{equation}
\label{eq:NiceIntOp}
\mathcal{K}[f](L)=\mathcal{F}(L)=\frac{k_{c}^{2}}{\pi}~\text{Im}\int_{0}^{
\infty} f(ix)~(1-2k_cLx)~e^{-2k_cLx}\,dx
\end{equation}
where Im stands for taking the imaginary part. To verify that the action
of this operator on all polynomial $f$ agrees with that given in
(\ref{dop}), let us begin by introducing variables $\Lambda=2k_{c}L$ and
$\tilde{x}=2k_cLx$, to write:
\begin{equation}
\label{eq:IntOpBeg} \mathcal{F}(L)=\frac{k_{c}^{2}}{\pi\Lambda}~\text{Im}
\int_{0}^{\infty} f\left(i\frac{\tilde{x}}{\Lambda}\right)(1-\tilde{x})~
e^{-\tilde{x}}\,d\tilde{x}
\end{equation}
We claim that iterated integrations by parts yield:
\begin{eqnarray} \label{eq:IntOp} \mathcal{F}(L)
&=&\frac{k_{c}^{2}}{\pi\Lambda}~\text{Im}\left\{
\sum_{s=0}^{n}\left.e^{-\tilde{x}}(\tilde{x}+s)\frac{d^{s}}{
d\tilde{x}^{s}}f\left(i\frac{\tilde{x}}{\Lambda}
\right)\right|_{\tilde{x}=0}^{ \infty}\right.
\nonumber \\
& & \qquad \quad \left. - \int_{0}^{\infty}e^{-\tilde{x}}(\tilde{x}+n)
\frac{d^{n+1}}{d\tilde{x}^{n+1}}f \left(i\frac{\tilde{x}}{\Lambda}\right)
\right\}
\end{eqnarray}
Integrating (\ref{eq:IntOpBeg}) by parts once shows that the equation
holds for $n=0$. Assuming now that the formula is valid for $n-1$,
integration by parts of the remaining integral yields:
\begin{eqnarray}
\mathcal{F}(L) &=& \frac{k_{c}^{2}}{\pi\Lambda}~\text{Im}\left\{
\sum_{s=0}^{n-1}\left.e^{-\tilde{x}}(x+s)\frac{d^{s}}{
d\tilde{x}^{s}}f\left(i\frac{\tilde{x}}{
\Lambda}\right)\right|_{\tilde{x}=0}^{ \infty}\right.
\label{wed1} \\
& & \qquad \qquad +
\left.e^{-\tilde{x}}(\tilde{x}+n)\frac{d^{n}}{d\tilde{x}^{n}}f\left(
i\frac{\tilde{x}}{\Lambda}\right)
\right|_{\tilde{x}=0}^{\infty}\label{wed2}\\
 &  & \qquad \qquad - \left. \int_{0}^{ \infty}
 e^{-\tilde{x}}(\tilde{x}+n)\frac{d^{n+1}}{d\tilde{x}^{n+1}}f\left(i
\frac{\tilde{x}}{\Lambda}\right)\right\}
\end{eqnarray}
The boundary term in (\ref{wed2}) becomes the next term in the sum
(\ref{wed1}) and by induction this completes the proof of
(\ref{eq:IntOp}). In (\ref{eq:IntOp}), since $f$ is polynomial, the
integral vanishes if $n$ is chosen large enough. Also, the boundary terms
clearly vanish at the upper limit. Letting $n\to \infty$, we are left
with:
\begin{eqnarray}
\label{eq:SumI} \mathcal{F}(L) & = &
\frac{-k_c^2}{\pi\Lambda}~\text{Im}\sum_{s=0}^\infty
\left.s\frac{d^{s}}{d\tilde{x}^{s}}f
\left(i\frac{\tilde{x}}{\Lambda}\right)\right|_{\tilde{x}=0}\\
 & = & \frac{k_c}{2\pi L} \label{la2}
\sum_{r=1}^\infty ~(2r-1)~(-1)^r~\left(\frac{1}{2k_cL}
\right)^{2r-1}\frac{d{\,}^{2r-1}}{dx^{2r-1}} ~f(x)\vert_{x=0}
\end{eqnarray}
which agrees with (\ref{dop}), up to the zeta function which we omitted
since it is close to one. In the step from (\ref{eq:SumI}) to (\ref{la2})
we made use of the fact that the imaginary part selects for only the odd
powers in the series.
\smallskip\newline
As a consistency check, let us apply the integral representation,
(\ref{eq:NiceIntOp}), of $\mathcal{K}$ to the usual linear dispersion
relation $f(x)=x$. Carrying out the integration yields
$\mathcal{F}(L)=-\frac{1}{4\pi L^{2}}$. As expected, this differs from the
usual result only by the omitted $\zeta$ function pre-factor of
$\zeta(2)=\frac{\pi^{2}}{6}$.

\subsection{Relation of $\mathcal{K}$ to the Laplace transform}

The representation of $\mathcal{K}$ as an integral operator came at the
cost of complexifying the analysis by having to integrate the dispersion
relation along the imaginary axis.

Fortunately, it is possible to re-express $\mathcal{K}$ as a real integral
operator, namely as a slightly modified Laplace transform. To this end,
let us use our finding that even powers in the dispersion relations do not
contribute to the Casimir force. This means that, without restricting
generality, we can assume that the dispersion relation is odd, i.e. that
it can be written in the form
\begin{equation}
f(x)=x~g(x^{2})
\end{equation}
for some function $g$. Thus, $f(ix) = i\,x\,g(-x^2)$, and therefore the
integral representation (\ref{eq:NiceIntOp}) of $\mathcal{K}$ now takes
the form:
\begin{equation}
\label{eq:OpIntRe}
\mathcal{K}[f](L)=\mathcal{F}(L)=\frac{k_{c}^{2}}{\pi}\int_{0}^{
\infty}x~g(-x^{2})~(1-2k_cLx)~e^{-2k_cLx}\,dx
\end{equation}
Using the properties of the Laplace transform with respect to
differentiation, we can finally conclude that the operator $\mathcal{K}$
which maps dispersion relations into Casimir force functions can be
written as a modified Laplace transform:
\begin{eqnarray}
\mathcal{K}[f](L) = \mathcal{F}(L) & = & \frac{k_{c}^{2}}{\pi}\nonumber
\left(1+L\frac{d}{d L}\right) \int_0^\infty e^{-2k_cLx}x~g(-x^2)~dx
\\
 & = & \frac{k_{c}^{2}}{\pi}
\left(1+L\frac{d}{d L}\right)\mathcal{L}_{\Lambda}[\tilde{f}]\label{cffi}
\end{eqnarray}
In the last line, $\mathcal{L}_{\Lambda}[\tilde{f}]$ stands for the
Laplace transform of $\tilde{f}(x) =x\,g(-x^{2})$ with respect to the
variable $\Lambda = 2k_cL$. Let us test (\ref{cffi}) by applying it to the
linear dispersion relation, where $\tilde{f}(x)=x$. Then,
\begin{eqnarray}
\mathcal{F}(L) & = & \frac{k_c^2}{\pi}~\left(1+L~\frac{d}{d
L}\right)\int_0^\infty
e^{-2k_c L x}x~dx\\
 & = & -\frac{1}{4\pi L^2}~,
\end{eqnarray}
which indeed agrees with the expected result as obtained at the end of
Sec.\ref{sec9.1}.
\smallskip\newline
We notice that the representation of $\mathcal{K}$ through (\ref{cffi})
involves the analytic extension of the function $g$ from positive
arguments, where it encodes the dispersion relation through
$f(x)=x\,g(x^2)$, to negative arguments where $g$ is evaluated by the
Laplace transform in (\ref{cffi}).
\smallskip\newline
This observation about $\mathcal{K}$ will be useful for answering the
transplanckian question in Sec.\ref{tps}: clearly, the dispersion relation
$f(x)=x\,g(x^2)$ may be very close to linear, i.e. $g(y)$ may be close to
one for $y>0$, while at the same time the unique analytic extension $g(y)$
for $y<0$ may be far from linear. This already shows that
ultraviolet-modified dispersion relations can easily lead to arbitrarily
pronounced nontrivial Casimir forces even at infrared length scales.

\subsection{The inverse of $\mathcal{K}$}

Let us now calculate the inverse of the operator $\mathcal{K}$ to obtain
the operator which maps odd Casimir force functions (recall that the even
ones do not contribute to the Casimir force) into the corresponding
dispersion relations. To this end, we need to solve for
$\tilde{\mathcal{F}}(L)$:
\begin{equation}
\frac{k_{c}^{2}}{\pi}\left(1+L\frac{d}{dL}\right)
\tilde{\mathcal{F}}(L)=\mathcal{F}(L)\,.
\end{equation}
The Green's function for this differential operator satisfies the
following equation:
\begin{equation}
\frac{k_{c}^{2}}{\pi}\left(1+L\frac{d}{dL}\right)
G_{\mathcal{F}}(L,L')=\delta(L-L')
\end{equation}
Since the $\delta$-function is formally the derivative of the Heavyside
step function $\theta$, an integration on both sides yields
\begin{equation}
\int G_{\mathcal{F}}(L,L')\,dL+L G_{\mathcal{F}}(L,L')-\int
G_{\mathcal{F}}(L,L')\,dL
=\frac{\pi}{k_{c}^{2}}\,\theta(L-L')+\kappa(L')\,,
\end{equation}
where $\kappa(L')$ is some arbitrary function. Hence,
\begin{equation}
G_{\mathcal{F}}(L,L')=\frac{1}{L}\left[\frac{\pi}{
k_{c}^{2}}\,\theta(L-L')+\kappa(L')\right]\,,
\end{equation}
and
\begin{equation}
\label{eq:PreFTilde}
\tilde{\mathcal{F}}(L)=\frac{1}{L}\int_{-\infty}^{\infty}
\left[\frac{\pi}{k_{c}^{2}}\,\theta(L-L')+
\kappa(L')\right]\mathcal{F}(L')\,dL'\,.
\end{equation}
For the boundary condition, we set $\tilde{\mathcal{F}}(L)\to0$ as $
L\to+\infty$, to ensure the correct behavior of $\mathcal{F}$. Hence,
\begin{equation}
\kappa(L')+\frac{\pi}{k_{c}^{2}}=0 \, \Longleftrightarrow \,
\kappa(L')\equiv -\frac{\pi}{k_{c}^{2}}
\end{equation}
Thus, the integral in (\ref{eq:PreFTilde}) is effectively truncated and we
have:
\begin{equation}
\label{eq:FTilde} \tilde{\mathcal{F}}(L)=-\frac{\pi}{k_{c}^{2}L}\int_{L}^{
\infty}\mathcal{F}(L')\,dL'\,.
\end{equation}
Eventually, we also need to invert the Laplace transform through a
Fourier-Mellin integral, to obtain:
\begin{equation}
\label{invfo} x\,g(-x^{2})=-\frac{1}{2i k_c^2L}\int_{\gamma}dL\,e^{x
L}\int_{L}^{\infty}\mathcal{F}(L')\,dL'
\end{equation}
Here, the integration path $\gamma$ is to be chosen parallel to the
imaginary axis and to the right of all singularities of the integrand.
Analytic continuation of $g$ to the positive reals finally yields the
dispersion relation $\mathcal{K}^{-1}[\mathcal{F}](x)=f(x)=xg(x^2)$,
modulo, of course, even components to the dispersion relations. We will
here not go further into the functional analysis of (\ref{invfo}) and the
inverse of $\mathcal{K}$.

\section{The transplanckian question}
\label{tps} Having calculated $\mathcal{K}$, we are now prepared to
address the transplanckian question, namely the question which types of
Planck scale modified dispersion relations would significantly affect the
predictions for the Casimir force at realistic plate separations.

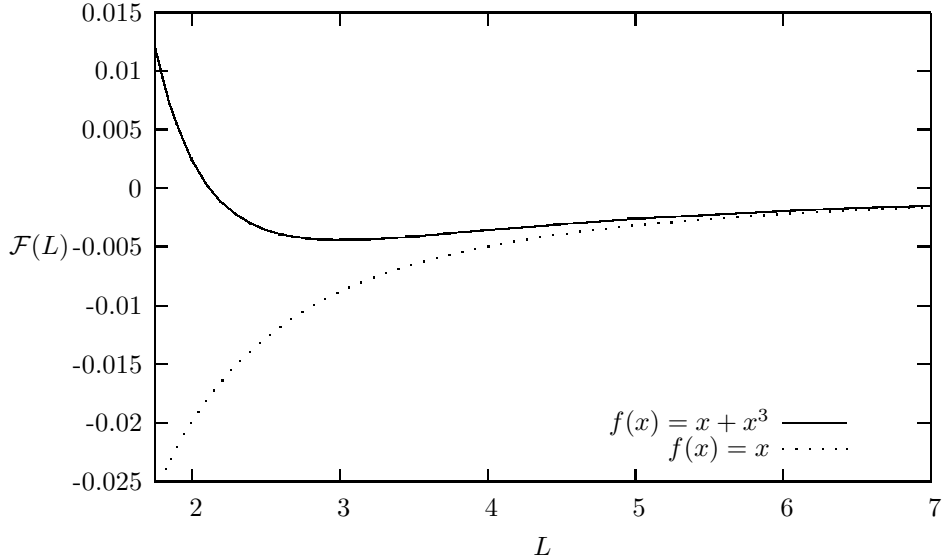
\begin{figure}[!ht]
\begin{center}
\setlength{\unitlength}{0.240900pt}
\ifx\plotpoint\undefined\newsavebox{\plotpoint}\fi
\begin{picture}(1500,900)(0,0)
\sbox{\plotpoint}{\rule[-0.200pt]{0.400pt}{0.400pt}}%
\put(221.0,123.0){\rule[-0.200pt]{4.818pt}{0.400pt}}
\put(201,123){\makebox(0,0)[r]{-0.025}}
\put(1419.0,123.0){\rule[-0.200pt]{4.818pt}{0.400pt}}
\put(221.0,215.0){\rule[-0.200pt]{4.818pt}{0.400pt}}
\put(201,215){\makebox(0,0)[r]{-0.02}}
\put(1419.0,215.0){\rule[-0.200pt]{4.818pt}{0.400pt}}
\put(221.0,307.0){\rule[-0.200pt]{4.818pt}{0.400pt}}
\put(201,307){\makebox(0,0)[r]{-0.015}}
\put(1419.0,307.0){\rule[-0.200pt]{4.818pt}{0.400pt}}
\put(221.0,399.0){\rule[-0.200pt]{4.818pt}{0.400pt}}
\put(201,399){\makebox(0,0)[r]{-0.01}}
\put(1419.0,399.0){\rule[-0.200pt]{4.818pt}{0.400pt}}
\put(221.0,492.0){\rule[-0.200pt]{4.818pt}{0.400pt}}
\put(201,492){\makebox(0,0)[r]{-0.005}}
\put(1419.0,492.0){\rule[-0.200pt]{4.818pt}{0.400pt}}
\put(221.0,584.0){\rule[-0.200pt]{4.818pt}{0.400pt}}
\put(201,584){\makebox(0,0)[r]{ 0}}
\put(1419.0,584.0){\rule[-0.200pt]{4.818pt}{0.400pt}}
\put(221.0,676.0){\rule[-0.200pt]{4.818pt}{0.400pt}}
\put(201,676){\makebox(0,0)[r]{ 0.005}}
\put(1419.0,676.0){\rule[-0.200pt]{4.818pt}{0.400pt}}
\put(221.0,768.0){\rule[-0.200pt]{4.818pt}{0.400pt}}
\put(201,768){\makebox(0,0)[r]{ 0.01}}
\put(1419.0,768.0){\rule[-0.200pt]{4.818pt}{0.400pt}}
\put(221.0,860.0){\rule[-0.200pt]{4.818pt}{0.400pt}}
\put(201,860){\makebox(0,0)[r]{ 0.015}}
\put(1419.0,860.0){\rule[-0.200pt]{4.818pt}{0.400pt}}
\put(279.0,123.0){\rule[-0.200pt]{0.400pt}{4.818pt}}
\put(279,82){\makebox(0,0){ 2}}
\put(279.0,840.0){\rule[-0.200pt]{0.400pt}{4.818pt}}
\put(511.0,123.0){\rule[-0.200pt]{0.400pt}{4.818pt}}
\put(511,82){\makebox(0,0){ 3}}
\put(511.0,840.0){\rule[-0.200pt]{0.400pt}{4.818pt}}
\put(743.0,123.0){\rule[-0.200pt]{0.400pt}{4.818pt}}
\put(743,82){\makebox(0,0){ 4}}
\put(743.0,840.0){\rule[-0.200pt]{0.400pt}{4.818pt}}
\put(975.0,123.0){\rule[-0.200pt]{0.400pt}{4.818pt}}
\put(975,82){\makebox(0,0){ 5}}
\put(975.0,840.0){\rule[-0.200pt]{0.400pt}{4.818pt}}
\put(1207.0,123.0){\rule[-0.200pt]{0.400pt}{4.818pt}}
\put(1207,82){\makebox(0,0){ 6}}
\put(1207.0,840.0){\rule[-0.200pt]{0.400pt}{4.818pt}}
\put(1439.0,123.0){\rule[-0.200pt]{0.400pt}{4.818pt}}
\put(1439,82){\makebox(0,0){ 7}}
\put(1439.0,840.0){\rule[-0.200pt]{0.400pt}{4.818pt}}
\put(221.0,123.0){\rule[-0.200pt]{293.416pt}{0.400pt}}
\put(1439.0,123.0){\rule[-0.200pt]{0.400pt}{177.543pt}}
\put(221.0,860.0){\rule[-0.200pt]{293.416pt}{0.400pt}}
\put(221.0,123.0){\rule[-0.200pt]{0.400pt}{177.543pt}}
\put(40,491){\makebox(0,0){$\mathcal{F}(L)$}}
\put(830,21){\makebox(0,0){$L$}}
\put(1187,215){\makebox(0,0)[r]{$f(x)=x+x^3$}}
\put(1207.0,215.0){\rule[-0.200pt]{24.090pt}{0.400pt}}
\put(221,808){\usebox{\plotpoint}}
\multiput(221.58,800.94)(0.492,-2.047){21}{\rule{0.119pt}{1.700pt}}
\multiput(220.17,804.47)(12.000,-44.472){2}{\rule{0.400pt}{0.850pt}}
\multiput(233.58,753.25)(0.492,-1.958){19}{\rule{0.118pt}{1.627pt}}
\multiput(232.17,756.62)(11.000,-38.623){2}{\rule{0.400pt}{0.814pt}}
\multiput(244.58,712.88)(0.492,-1.444){21}{\rule{0.119pt}{1.233pt}}
\multiput(243.17,715.44)(12.000,-31.440){2}{\rule{0.400pt}{0.617pt}}
\multiput(256.58,679.61)(0.496,-1.205){43}{\rule{0.120pt}{1.057pt}}
\multiput(255.17,681.81)(23.000,-52.807){2}{\rule{0.400pt}{0.528pt}}
\multiput(279.58,625.77)(0.496,-0.852){43}{\rule{0.120pt}{0.778pt}}
\multiput(278.17,627.38)(23.000,-37.385){2}{\rule{0.400pt}{0.389pt}}
\multiput(302.58,587.56)(0.496,-0.609){43}{\rule{0.120pt}{0.587pt}}
\multiput(301.17,588.78)(23.000,-26.782){2}{\rule{0.400pt}{0.293pt}}
\multiput(325.00,560.92)(0.600,-0.496){37}{\rule{0.580pt}{0.119pt}}
\multiput(325.00,561.17)(22.796,-20.000){2}{\rule{0.290pt}{0.400pt}}
\multiput(349.00,540.92)(0.827,-0.494){25}{\rule{0.757pt}{0.119pt}}
\multiput(349.00,541.17)(21.429,-14.000){2}{\rule{0.379pt}{0.400pt}}
\multiput(372.00,526.92)(1.173,-0.491){17}{\rule{1.020pt}{0.118pt}}
\multiput(372.00,527.17)(20.883,-10.000){2}{\rule{0.510pt}{0.400pt}}
\multiput(395.00,516.93)(1.713,-0.485){11}{\rule{1.414pt}{0.117pt}}
\multiput(395.00,517.17)(20.065,-7.000){2}{\rule{0.707pt}{0.400pt}}
\multiput(418.00,509.94)(3.259,-0.468){5}{\rule{2.400pt}{0.113pt}}
\multiput(418.00,510.17)(18.019,-4.000){2}{\rule{1.200pt}{0.400pt}}
\multiput(441.00,505.95)(5.151,-0.447){3}{\rule{3.300pt}{0.108pt}}
\multiput(441.00,506.17)(17.151,-3.000){2}{\rule{1.650pt}{0.400pt}}
\put(465,502.67){\rule{5.541pt}{0.400pt}}
\multiput(465.00,503.17)(11.500,-1.000){2}{\rule{2.770pt}{0.400pt}}
\put(488,501.67){\rule{5.541pt}{0.400pt}}
\multiput(488.00,502.17)(11.500,-1.000){2}{\rule{2.770pt}{0.400pt}}
\put(511,502.17){\rule{11.700pt}{0.400pt}}
\multiput(511.00,501.17)(33.716,2.000){2}{\rule{5.850pt}{0.400pt}}
\multiput(569.00,504.60)(8.377,0.468){5}{\rule{5.900pt}{0.113pt}}
\multiput(569.00,503.17)(45.754,4.000){2}{\rule{2.950pt}{0.400pt}}
\multiput(627.00,508.59)(6.387,0.477){7}{\rule{4.740pt}{0.115pt}}
\multiput(627.00,507.17)(48.162,5.000){2}{\rule{2.370pt}{0.400pt}}
\multiput(685.00,513.59)(6.387,0.477){7}{\rule{4.740pt}{0.115pt}}
\multiput(685.00,512.17)(48.162,5.000){2}{\rule{2.370pt}{0.400pt}}
\multiput(743.00,518.59)(6.727,0.489){15}{\rule{5.256pt}{0.118pt}}
\multiput(743.00,517.17)(105.092,9.000){2}{\rule{2.628pt}{0.400pt}}
\multiput(859.00,527.59)(6.727,0.489){15}{\rule{5.256pt}{0.118pt}}
\multiput(859.00,526.17)(105.092,9.000){2}{\rule{2.628pt}{0.400pt}}
\multiput(975.00,536.59)(10.435,0.482){9}{\rule{7.833pt}{0.116pt}}
\multiput(975.00,535.17)(99.742,6.000){2}{\rule{3.917pt}{0.400pt}}
\multiput(1091.00,542.59)(10.435,0.482){9}{\rule{7.833pt}{0.116pt}}
\multiput(1091.00,541.17)(99.742,6.000){2}{\rule{3.917pt}{0.400pt}}
\multiput(1207.00,548.59)(12.844,0.477){7}{\rule{9.380pt}{0.115pt}}
\multiput(1207.00,547.17)(96.531,5.000){2}{\rule{4.690pt}{0.400pt}}
\multiput(1323.00,553.61)(25.691,0.447){3}{\rule{15.567pt}{0.108pt}}
\multiput(1323.00,552.17)(83.691,3.000){2}{\rule{7.783pt}{0.400pt}}
\put(1187,174){\makebox(0,0)[r]{$f(x)=x$}}
\multiput(1207,174)(20.756,0.000){5}{\usebox{\plotpoint}}
\put(1307,174){\usebox{\plotpoint}}
\put(229.00,123.00){\usebox{\plotpoint}}
\put(237.61,141.86){\usebox{\plotpoint}}
\multiput(246,158)(9.601,18.402){2}{\usebox{\plotpoint}}
\put(266.68,196.92){\usebox{\plotpoint}}
\put(277.54,214.59){\usebox{\plotpoint}}
\put(288.95,231.93){\usebox{\plotpoint}}
\put(300.68,249.05){\usebox{\plotpoint}}
\put(313.12,265.65){\usebox{\plotpoint}}
\put(326.43,281.57){\usebox{\plotpoint}}
\put(339.97,297.30){\usebox{\plotpoint}}
\put(354.30,312.30){\usebox{\plotpoint}}
\put(369.49,326.45){\usebox{\plotpoint}}
\put(384.79,340.47){\usebox{\plotpoint}}
\put(400.90,353.47){\usebox{\plotpoint}}
\put(417.65,365.73){\usebox{\plotpoint}}
\put(434.42,377.95){\usebox{\plotpoint}}
\put(451.92,389.10){\usebox{\plotpoint}}
\put(469.80,399.63){\usebox{\plotpoint}}
\put(488.18,409.24){\usebox{\plotpoint}}
\put(506.80,418.40){\usebox{\plotpoint}}
\put(525.50,427.39){\usebox{\plotpoint}}
\put(544.61,435.50){\usebox{\plotpoint}}
\put(563.76,443.49){\usebox{\plotpoint}}
\put(583.52,449.84){\usebox{\plotpoint}}
\put(603.22,456.37){\usebox{\plotpoint}}
\put(622.99,462.66){\usebox{\plotpoint}}
\put(642.98,468.22){\usebox{\plotpoint}}
\put(662.98,473.74){\usebox{\plotpoint}}
\put(683.15,478.65){\usebox{\plotpoint}}
\put(703.31,483.58){\usebox{\plotpoint}}
\put(723.62,487.77){\usebox{\plotpoint}}
\put(743.93,491.99){\usebox{\plotpoint}}
\put(764.21,496.34){\usebox{\plotpoint}}
\put(784.70,499.62){\usebox{\plotpoint}}
\put(805.19,502.95){\usebox{\plotpoint}}
\put(825.68,506.28){\usebox{\plotpoint}}
\put(846.15,509.69){\usebox{\plotpoint}}
\put(866.76,511.96){\usebox{\plotpoint}}
\put(887.26,515.17){\usebox{\plotpoint}}
\put(907.85,517.64){\usebox{\plotpoint}}
\put(928.45,519.99){\usebox{\plotpoint}}
\put(949.08,522.17){\usebox{\plotpoint}}
\put(969.66,524.78){\usebox{\plotpoint}}
\put(990.33,526.53){\usebox{\plotpoint}}
\put(1011.02,528.23){\usebox{\plotpoint}}
\put(1031.71,529.89){\usebox{\plotpoint}}
\put(1052.33,532.13){\usebox{\plotpoint}}
\put(1072.98,534.00){\usebox{\plotpoint}}
\put(1093.70,534.90){\usebox{\plotpoint}}
\put(1114.38,536.62){\usebox{\plotpoint}}
\put(1135.07,538.31){\usebox{\plotpoint}}
\put(1155.76,539.98){\usebox{\plotpoint}}
\put(1176.47,541.00){\usebox{\plotpoint}}
\put(1197.17,542.35){\usebox{\plotpoint}}
\put(1217.86,543.99){\usebox{\plotpoint}}
\put(1238.59,544.72){\usebox{\plotpoint}}
\put(1259.29,546.00){\usebox{\plotpoint}}
\put(1280.00,547.00){\usebox{\plotpoint}}
\put(1300.73,547.75){\usebox{\plotpoint}}
\put(1321.43,549.00){\usebox{\plotpoint}}
\put(1342.15,550.00){\usebox{\plotpoint}}
\put(1362.87,550.82){\usebox{\plotpoint}}
\put(1383.60,551.51){\usebox{\plotpoint}}
\put(1404.33,552.19){\usebox{\plotpoint}}
\put(1425.05,553.00){\usebox{\plotpoint}}
\put(1439,554){\usebox{\plotpoint}}
\put(221.0,123.0){\rule[-0.200pt]{293.416pt}{0.400pt}}
\put(1439.0,123.0){\rule[-0.200pt]{0.400pt}{177.543pt}}
\put(221.0,860.0){\rule[-0.200pt]{293.416pt}{0.400pt}}
\put(221.0,123.0){\rule[-0.200pt]{0.400pt}{177.543pt}}
\end{picture}

\caption{The Casimir force for a lowest order correction to the dispersion
relation: $f(x)=x+x^3$. The plate separation, $L$, is measured in
multiples of the UV scale $\frac{k_c^{-1}}{2}$.} \label{fig:Poly}
\end{center}
\end {figure}

To this end, let us begin by investigating the lowest order corrections to
the dispersion relation, $f$, namely by including a quadratic and a
quartic correction term: $f(x)=x+\nu_2x^2+\nu_3x^3$. The coefficients
$\nu_2,\nu_3$ can be as large as of order one, $\nu_2,\nu_3\approx 1$,
without appreciably affecting the dispersion relation
$\omega(k)=k_c\,f(k/k_c)$ at small momenta $k\ll k_c$. Using our result
(\ref{eq:forceFactorial}) for $\mathcal{K}$ we find the corresponding
Casimir force function:
\begin{equation}
\mathcal{F}(L) = -\frac{\pi}{24 L^2} \,+\nu_3\,\frac{\pi^5}{20\, k_c^2L^4}
\end{equation}
The quadratic correction term $\nu_2 x^2$ is an even component of $f$ and
therefore does not affect the Casimir force. The quartic correction term
does affect the Casimir force, changing the Casimir force from attractive
to repulsive at very short distances, as shown in Fig.~\ref{fig:Poly}.
However, as we can also see in Fig.~\ref{fig:Poly}, the Casimir force
function converges very rapidly towards the usual Casimir force function
for plate separations that are significantly larger than $L_c=k_c^{-1}$.
To be precise, we recall that the standard dispersion relation
$f_\text{standard}(x)=x$ implies the standard Casimir force function
$\mathcal{F}_\text{standard}(L) = -\frac{\pi}{24 L^2}$. The relative size
of the correction to the Casimir force depends on the plate separation $L$
and reads:
\begin{equation}
\label{relcorc} \frac{\mathcal{F}_\text{standard}(L)-\mathcal{F}(L)}{
\mathcal{F}_\text{standard}(L)}=\nu_3\frac{6\pi^4}{5L^2k_c^2}
\end{equation}
Let us calculate the orders of magnitude. The dispersion relation
$\omega(k)=k_c\,f(k/k_c)$ is expected to start to appreciably differ from
linearity the latest at the Planck scale, which in $3+1$ dimensional
space-time means that the critical length, $L_c$, obeys $L_c
=k_{c}^{-1}\approx 10^{-35}m$. Actual measurements of the Casimir force
have been performed at about $L_m\approx 10^{-7}m$, see e.g.
\cite{Lamoreaux:1999cu-etal}. Therefore, evaluating the relative
correction of the Casimir force, (\ref{relcorc}), at the measurable scale
$L=L_m$ yields
\begin{equation}
\frac{\mathcal{F}_\text{standard}(L_m)-\mathcal{F}(L_m)}{
\mathcal{F}_\text{standard}(L_m)}=\nu_3\,\frac{6\pi^4}{5}~\sigma^2
\end{equation}
where $\sigma$ denotes the dimensionless ratio of the ultraviolet length
scale $L_c$ and the infrared length scale $L_m$:
\begin{equation}
\sigma=\frac{L_{c}}{L_{m}} \approx 10^{-28}
\end{equation}
Thus, the effect of the lowest order corrections to the dispersion
relation on the Casimir force is extremely small at measurable plate
separations.

Naively, on might expect that higher-order corrections to the dispersion
relations contribute even less to the Casimir force. If true, this would
indicate that the physical processes that happen at these two length
scales respectively are very effectively decoupled from another. In fact,
however, the two scales are not quite as decoupled. Roughly speaking, the
reason is that higher order corrections to the dispersion relations
contribute more rather than less to the Casimir force, as we will now
show.

\subsection{UV-IR coupling with polynomial dispersion relations}
\label{sec:IRUV} \label{shfi} Recall that we here need not be concerned
with the even components of dispersion relations since they do not
contribute to the Casimir force. Let us, therefore, consider higher order
odd polynomial dispersion relations:
\begin{equation}
f(x)=x+\sum_{r=2}^N\nu_{2r-1}\,x^{2r-1}
\end{equation}
The coefficients $\nu_{2r-1}$ can be chosen as large as of order one,
$\nu_{2r-1}\approx 1$, and $f$ will still be modified only in the
ultraviolet. We showed above that the contribution of the lowest order
correction term, $\nu_3x^3$, to the Casimir force at the infrared length
scale $L_m$ is proportional to $\sigma^2$, i.e. that it is completely
negligible. One might expect that higher order terms $\nu_{2r-1}x^{2r-1}$
in the dispersion relation would contribute even less to the Casimir
force. At first sight this expectation appears to be confirmed:
$\mathcal{K}$ maps a dispersion relation term $\sim x^{2n-1}$ into a
Casimir force term $\sim (k_cL)^{-2r}$. At the infrared scale, $L=L_m$,
the latter term reads:
\begin{equation}
\left(\frac{1}{k_cL}\right)^{2r}=\left(\frac{L_c}{L_m}\right)^{2r}=
\sigma^{2r}
\end{equation}
This indeed means that the size of this term decreases exponentially with
increasing $r$. Upon closer inspection, however, we see that,
nevertheless, a higher order term $x^{2r-1}$ in $f$ can give an
arbitrarily large contribution to the Casimir force, in particular if $r$
is very large. The reason is that $\mathcal{K}$ involves a factorial
amplification of higher order terms which eventually overcomes the
exponential suppression that we discussed above. Namely, as
(\ref{eq:forceFactorial}) shows, the precise action of $\mathcal{K}$ on
the correction term $\nu_{2r-1}x^{2r-1}$ reads:
\begin{equation}
\mathcal{K}: ~~\nu_{2r-1}\,x^{2r-1} ~~\longrightarrow~~
\nu_{2r-1}\,\frac{(-1)^r k_c^2}{\pi}\,(2r-1)(2r-1)!\,
\zeta(2r)\left(\frac{1}{2k_cL}\right)^{2r} \label{kappa34}
\end{equation}
Due to the presence of the factorial term $(2r-1)!$, the coefficients of
the Casimir force function grow much faster than those of the dispersion
relation. In particular, for the dispersion relation $f(x)=x+
\nu_{2r-1}x^{2r-1}$ the relative change in the Casimir force at the
infrared scale $L_m$ reads:
\begin{equation}
\frac{\mathcal{F}_\text{standard}(L_m)-\mathcal{F}(L_m)}{
\mathcal{F}_\text{standard}(L_m)}=\nu_{2r-1}
\frac{(-1)^{r-1}(2r-1)\zeta(2r)}{4\pi^2}~(2r-1)!~
\left(\frac{\sigma}{2}\right)^{2r-2}
\end{equation}
It is straightforward to apply Stirling's formula for the factorial, $
n!\approx \sqrt{2\pi n}\left(\frac{n}{e}\right)^{n}$ for $n\gg 1$ in order
to calculate how large $r$ needs to be for the factorial amplification to
overcome the exponential suppression. We find that a correction term
$\nu_{2r-1}x^{2r-1}$ with $\nu_{2r-1}\approx1$ in the dispersion relation
leads to a relative change of order one in the Casimir force at the
infrared scale $L_m$ if $r$ is of the order $\sigma^{-1}$, i.e. if
$r\approx 10^{28}$.
\smallskip\newline
To summarize: We found that $\mathcal{K}$ is a well-defined but unbounded
and therefore discontinuous operator (as are, e.g., the quantum mechanical
position and momentum operators). Namely, a modified dispersion relation
of the form $f(x) = x + \nu_{2r-1}x^{2r-1}$, say with $r\approx10^{28}$
and $\nu_{2r-1}\approx 1$ is virtually indistinguishable from the linear
dispersion relation $f(x)=x$ at all scales up to the Planck scale, but
does lead to a modification of the Casimir force which is very strong (the
relative change is of order 100\%) even at laboratory length scales. Thus,
even though the first order terms contribute extremely little to the
Casimir force, very high order corrections to the dispersion relations can
contribute significantly to the Casimir force - in fact, the more so the
larger $r$ is.

Realistic candidates for Planck scale modified dispersion relation are
given by a series $f(x)=x+\sum_{n=2}^\infty \nu_n x^n$ and such dispersion
relations therefore contain terms $\nu_{2r-1}x^{2r-1}$ for arbitrarily
large $r$. At the same time, the prefactors $\nu_n$ must of course obey
$\nu_n\rightarrow 0$ as $n\rightarrow \infty$ because this is a necessary
condition for the convergence of the series. We conclude that it is this
competition between the decay of the coefficients $\nu_{2r-1}$ and the
increasing Casimir effect of terms $x^{2r-1}$, for $r\rightarrow\infty$,
which decides whether or not a given ultraviolet-modified dispersion
relation does or does not lead to an appreciable effect on the Casimir
force at infrared distances. In practice, to study this competition
directly by using the complicated representation of $\mathcal{K}$ in
(\ref{kappa34}) would be a tedious approach to the transplanckian question
because, for example, the coefficients of the Casimir force acquire
alternating signs. Instead, as we will show in the next section, we will
conveniently be able to study the transplanckian question by making use of
our representation of $\mathcal{K}$ in terms of the Laplace transform.

\subsection{UV-IR coupling with generic dispersion
relations} \label{shsec}

Let us write the dispersion relations again in the form $f(x)=x\,g(x^2)$
so that, e.g., $g\equiv 1$ yields the standard dispersion relation. This
allows us to apply the representation of $\mathcal{K}$ in terms of the
Laplace transform, (\ref{eq:OpIntRe}). We begin by noticing that, since
$x^2$ is positive, the evaluation of the dispersion relation $f$ involves
evaluating $g(y)$ only for positive $y$. Now considering
(\ref{eq:OpIntRe}) we see that, curiously, the calculation of the Casimir
force involves evaluating $g(y)$ only for negative values of $y$.

This is surprising because if $g$ could be any arbitrary function, this
would mean that the dispersion relation, which is determined by the
behavior of $g$ on the positive half-axis, and the Casimir force function,
which is determined by the behavior of $g$ on the negative half axis, were
unrelated. But of course our $g$ are not arbitrary functions but are
polynomials or power series with infinite radius of convergence, i.e. they
are entire functions. Therefore, the behavior of $g$ on the positive half
axis fully determines its behavior also on the negative half axis. The
dispersion relations do determine the corresponding Casimir force.

\begin{figure}[!ht]
\begin{center}
\input{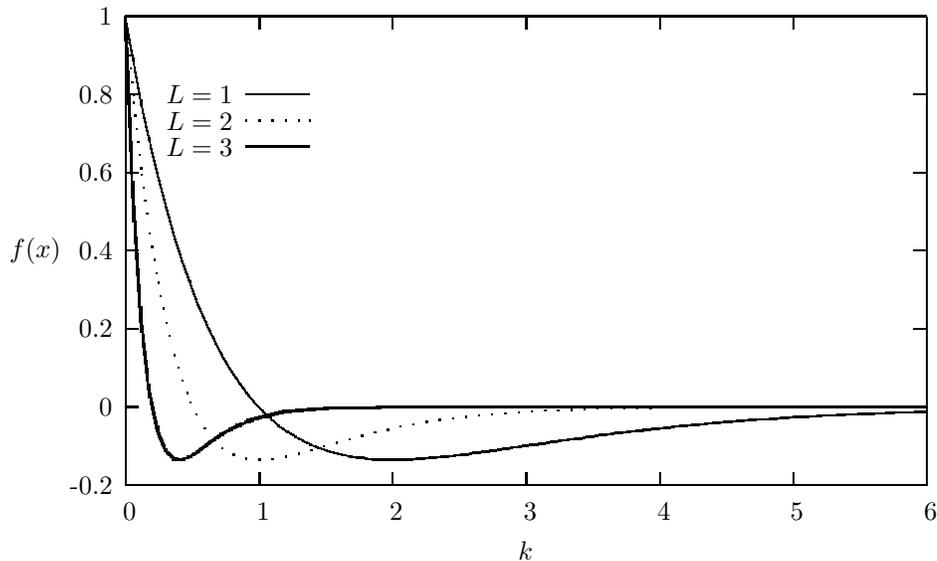}
\caption[The integral kernel]{The integral kernel $(1-xL)e^{-xL}$
 for different values of $L$. Note the shift of its zero towards
 the origin as $L$ grows.}
\label{fig:kernel}
\end{center}
\end {figure}

Of crucial importance for the transplanckian question, however, is the
fact that there are entire functions $g$ which are arbitrarily close to
one for $0<y<1$ and which nevertheless reach arbitrarily large values on
the negative half axis. Such functions do not noticeably affect the
dispersion relation for momenta up to the Planck scale but do arbitrarily
strongly affect the Casimir force. These are the dispersion relations
$f(x)=x\,g(x^2)$ with
\begin{equation}
g(y)=1+h(y),\label{def77}
\end{equation}
where the function $h$ obeys $h(y)\approx 0$ for $y\in(0,1)$ while
exhibiting large $\vert h(y)\vert$ in some range of negative values of
$y$. Let us now analyze which behavior of $h$ on the negative half axis
determines if the Casimir force is affected in the infrared. To this end,
let us use (\ref{eq:OpIntRe}) and (\ref{def77}) to express the correction
in the Casimir force, $\Delta
\mathcal{F}=\mathcal{F}-\mathcal{F}_\text{standard}$, in terms of the
correction $h$ to the dispersion relation:
\begin{equation}
\Delta\mathcal{F}(L)=\frac{k_{c}^{2}}{\pi}\int_{0}^{
\infty}x~h(-x^{2})~(1-2k_cLx)~e^{-2k_cLx}\,dx
\end{equation}
The integral kernel
\begin{equation}
G(x, L)=(1-2k_cLx)~e^{-2k_cLx}
\end{equation}
is positive for $x<(2k_cL)^{-1}$, negative for $x>(2k_cL)^{-1}$ and
rapidly decreases to zero for $x\gg(2k_cL)^{-1}$. (We remark that the the
integral of the kernel over all $x\in[0, \infty)$ is $0$, which expresses
the fact that the Casimir force does not depend on the absolute value of
the energy.) Thus, for a fixed plate separation $L$, what matters most for
the Casimir force is the behavior of $h(y)$ from $y=0$ to about
$y\approx-(k_cL)^{-2}$. As we increase $L$, the interval
$y\in(-(k_cL)^{-2},0)$ on which the integral kernel $G$ is mostly
supported is shrinking, see Fig.\ref{fig:kernel}.
Thus, there is a significant effect on the Casimir force at realistically
large plate separations, such as $L=L_m$, if the function $h$ is either of
order one in this small interval close to the origin or it must be
exponentially large (so as to compensate the exponential suppression in
$G$) in some interval to the left of $-(k_cL)^{-2}$. Of course, both are
possible. There are entire functions $h$ which possess either one of these
behaviors on the negative half axis and therefore do affect the Casimir
force in the infrared, while being arbitrarily close to zero for $0<y<1$,
so as to leave the dispersion relation virtually unchanged in the
infrared.

There is even the extreme case of functions, $h$, whose corresponding
dispersion relation $f$ is arbitrarily little affected at \it all \rm
scales while the Casimir force function is arbitrarily much affected at
any scale we wish, say in the infrared. To see this, consider for example
the case where $h$ is a Gaussian which is centred around a low negative
value $y_0<0$ while being so sharply peaked that its tail into the
positive half axis is negligibly small. The function that enters into the
calculation of the Casimir force, $\tilde{f}_1=x\,g(-x^2)$, then features
the low-$x$ spike of the Gaussian, implying by our above consideration
that the Casimir force is affected in the infrared. At the same time, the
dispersion relation itself, $\tilde{f}_2(x)=x\,g(x^2)$, is virtually
unaffected for all $x$.
\section{Conclusions}

We investigated the effect of ultraviolet corrections to the dispersion
relation on the Casimir force. To this end, we calculated the operator
$\mathcal{K}$ which maps generic dispersion relations,
$\omega(k)=k_{c}f\left(k/k_{c}\right)$, into the corresponding Casimir
force functions $\mathcal{F}(L)$. Here, $k_c$ is the Planck momentum, $f$
is a power series in $x=k/k_c$ and $L$ is the plate separation. The
structure of $\mathcal{K}$ showed that the even components of dispersion
relations do not contribute to the Casimir force. This implies, for
example, that the dispersion relations defined through $f(x)=\sinh(x)$ and
$f(x)=\exp(x)-1$ yield identical Casimir force functions.

We also showed that a certain class of UV-modified dispersion relations,
such as $f(x)=\sinh(x)$, lead to Casimir force functions that are well
defined only down to a finite smallest distance between the plates.
Physically, the existence of a finite lower bound for the plate
separation, $L$, is indeed what should be expected if the
ultraviolet-modified dispersion relation arises from an underlying theory
of quantum gravity which possesses a notion of minimum length.

Technically, the phenomenon of a finite minimum $L$ arises because the
Casimir force $\mathcal{F}(L)$ is always a polynomial or power series in
$1/L$, depending on whether the dispersion relation is polynomial or a
power series. Therefore, if $\mathcal{F}(L)$ is a power series then it can
possess a finite radius of convergence, i.e. an upper bound on $1/L$,
which then implies a lower bound on $L$. Of course, a finite radius of
convergence can occur only for power series but not for polynomials.
Interestingly, this means that the existence of a finite lower bound on
$L$ cannot arise from polynomial dispersion relations of any degree. An
important conclusion that we can draw from this is that if a candidate
quantum gravity theory yields a non-polynomial dispersion relation then
working with any finite degree polynomial approximation of this dispersion
relation may be missing crucial qualitative features, such as the
existence of a finite minimum length.

There is a deeper reason for why it is important to apply a nontrivial
dispersion relation in the exact form in which it arises from some
proposed quantum gravity theory. The reason is that $\mathcal{K}$ is an
unbounded and therefore also discontinuous operator, which means that
arbitrarily small changes to the dispersion relation can lead to
arbitrarily large changes to the Casimir force. On the other hand, the
action of $\mathcal{K}$ is of course well-defined, which means that if a
candidate quantum gravity theory implies a particular UV-modified
dispersion relation then $\mathcal{K}$ can be used to precisely predict
the corresponding Casimir force function.

We proceeded by determining which ultraviolet modifications to the
dispersion relation would appreciably affect the Casimir force function at
a large length scale $L_m$. To this end, it was convenient to express
dispersion relations, $f$, in the form $f(x)=x\,g(x^2)$ and $g(y)=1+h(y)$
where $h$ is an entire function (so that $h\equiv 0$ for the usual linear
dispersion relation). Recall that $y$ is the momentum squared, in units of
$k^2_c=L^{-2}_c$, i.e., $y=1$ is the Planck momentum squared. We are
interested in dispersion relations which are essentially unchanged in the
infrared, i.e., which obey $h(y)\approx 0$, up to unmeasurable deviations,
for all $y$ in the interval $(0,1)$. Our analysis of $\mathcal{K}$ through
the Laplace transform then showed that if the corresponding Casimir force
is to be affected at an infrared scale, say $L_m$, then the dispersion
relation must come from a function $h$ which obeys one or both of two
conditions: (a) either $h$ obeys $\vert h(y)\vert =\mathcal{O}(1)$ for $y$
in parts of the interval $(-L^2_c/L^2_m,0)=(-\sigma^2,0)$, or (b) $h$ is
exponentially large in a finite interval of more negative $y$ obeying
$y<-\sigma^2$.

In the case (a), an ultraviolet-modified dispersion relation induces an
infrared modification of the Casimir force if the correction to the
dispersion relation, $h(y)$, is essentially zero in all of $(0,1)$, while
it rises very steeply towards the left to amplitudes of order one within
the extremely short interval $(-\sigma^2,0)$, where we recall that
$\sigma\approx 10^{-28}$. In the case (b), UV/IR coupling arises if $h$ is
again essentially zero in the interval $(0,1)$, while now needing to reach
exponentially large values for a finite stretch of more negative $y$
values, again resulting in the need for $h$ to rise extremely steeply
towards the left. It is easy to give examples of such $h$, such as the
Gaussian $h$ that we discussed. In fact, we can easily write down $h$
which would lead to no appreciable modification of the dispersion at low
energies and yet to arbitrarily large changes to the Casimir force even at
macroscopically large plate separations. Because of their large slope,
however, such functions $h$ are severely fine-tuned and must therefore be
considered unlikely to arise from an underlying quantum gravity theory. We
can conclude, therefore, that the 28 orders of magnitude which separate
the effective UV and IR scales do not suppress UV/IR coupling in strength
but instead in likelihood, namely through the need for extreme fine
tuning.

This is interesting because, in inflation, the separation of the effective
UV and IR scales is only about three to five orders of magnitude: Consider
the operator $\mathcal{K}$ for inflation, namely the operator which maps
arbitrary ultraviolet-modified dispersion relations into the function that
describes the CMB's tensor or scalar fluctuation spectrum. Let us assume
that its properties are analogous to that of the operator $\mathcal{K}$
which we here found for the Casimir effect. This would mean that an
ultraviolet-modified dispersion relation that arises from some underlying
quantum gravity theory can lead to effects on the CMB spectrum which are
not automatically limited in their strength by the separation of scales
$\sigma\approx 10^{-5}$, or indeed by any power of $\sigma$. Instead,
arbitrarily large effects on the CMB must be considered possible, while it
is merely the a priori likelihood of large effects that is suppressed by
the separation of scales. That this is indeed the case can of course only
be confirmed by calculating an explicit expression for the operator
$\mathcal{K}$ for inflation.

\section{Outlook}

The task of finding the operator $\mathcal{K}$ for inflation will be more
difficult than it was to calculate $\mathcal{K}$ for the Casimir effect.
This is mainly because it is highly nontrivial to identify the comoving
modes' initial condition, i.e. their ingoing vacuum state. This problem
needs to be solved because a misidentification of the vacuum could mask
the infrared effects that one is looking for. The reason is that the mode
equations reduce to the mode equations with the usual linear dispersion at
late times, namely at large length scales. Therefore, the mode solutions
at late times live in the usual solution space. Thus, any effects of
ultraviolet-modified dispersion relations in the IR could be masked by an
incorrect choice of the initial condition for the mode equation. A further
complication is that of possibly strong backreaction, although there are
indications that this problem can be absorbed in a suitable redefinition
of the inflaton potential, see \cite{greenenew}. Once these points are
clarified, $\mathcal{K}$ for inflation can be calculated.

A limitation of our investigation of the Casimir effect has been that we
restricted attention to modelling the effects of Planck scale physics on
quantum field theory exclusively through UV-modified dispersion relations.
This assumes that fields can possess arbitrarily large $k$ and arbitrarily
short wavelengths, an assumption which is likely too strong. Indeed,
studies of quantum gravity and string theory strongly indicate the
existence of a universal minimum length at the Planck or string scale. In
particular, it has been suggested that, in terms of first quantization,
this natural UV cutoff could possess an effective description through
uncertainty relations of the form $\Delta x \Delta p \le
\frac{\hbar}{2}(1+\beta (\Delta p)^2 +...)$, see, e.g.,
\cite{Garay:1994en}. As is easily verified, such uncertainty relations
encode the minimum length as a lower bound, $\Delta
x_\text{min}=\hbar\sqrt{\beta}$, on the formal position uncertainty,
$\Delta x$. It has been shown that this type of uncertainty relations also
implies a minimum wavelength and that, therefore, fields possess the
sampling property, see \cite{ak-prl2000}: if a field's (number or
operator-valued) amplitudes are known only at discrete points then the
field's amplitudes everywhere are already determined - if the average
sample spacing is less than the critical spacing, which is given by the
minimum length. As a consequence, any theory with this type of uncertainty
relation can be written as continuum theory or, fully equivalently, as a
discrete theory on any lattice of sufficiently tight spacing. This UV
cutoff can also be viewed as an information theoretic cutoff, and it
possesses a covariant generalization, see \cite{ak-prl}.

Indeed, nontrivial dispersion relations also raise the question of local
Lorentz invariance. One possibility is that local Lorentz is broken hard
or soft and that, e.g., the CMB rest frame is the preferred frame. It has
also been suggested that the Lorentz group might be deformed, or that it
may be unchanged but represented nonlinearly. Various experimental bounds
on Lorentz symmetry breaking are being discussed, e.g., from observations
of gamma ray bursts. For the literature, see e.g. \cite{lorentz}.

An application of the minimum length uncertainty principle to the Casimir
effect has recently been tried, see \cite{hossen}. There, the Casimir
force was found to be a discontinuous function of the plate separation.
This problem is due to the fact that, in \cite{hossen}, the plate
boundaries are implicitly treated as possessing sharp positions. This is
not fully consistent with the assumption that all particles including
those that make up the plates can be localized only up to the finite
minimum position uncertainty. As a consequence, as the plate separation
increases, the energy eigenvalues discontinuously enter the spectrum  of
the first quantized Hamiltonian. It should be very interesting to extend
these Casimir force calculations while applying the minimum length
uncertainty relations to both the field and the plates.

Finally, we note an additional analogy between the Casimir effect and
inflation: in the Casimir effect with UV cutoff, as the distance between
the plates is increased, new modes enter the space between the plates,
thereby changing the vacuum energy. In cosmology, space itself expands
and, in the presence of an UV cutoff, new comoving modes (recall that
these are the independent degrees of freedom) are continually being
created, similar to the Casimir effect. A priori, these new modes arise
with vacuum energy. During the expansion, the modes' vacuum energy becomes
diluted but if the dispersion is nonlinear then the balance of new vacuum
energy creation and vacuum energy dilution is nontrivial. A paper which
addresses this question is in progress, \cite{ak-ll}.

\end{document}